\documentclass[twocolumn]{aastex631}
\usepackage{soul}
\usepackage{tikz}
\usepackage{hyperref}
\usepackage{subfigure}
\usepackage{graphicx}
	
\setcitestyle{notesep={}} 

\usepackage{float}
\usepackage{amsmath}
\newcommand{\gp}{\ensuremath{\gamma_+}}
\newcommand{\gx}{\ensuremath{\gamma_\times}}
\newcommand{\ATF}{\ensuremath{A_{\rm TF}}}
\newcommand{\vmax}{\ensuremath{v_{\rm max}}}
\usepackage{soul}

\submitjournal{ApJ}

\shorttitle{Velocity Field Lensing}

\graphicspath{{./}{figures/}}

\begin{document}
\title{Testing Velocity Field Lensing on IllustrisTNG Galaxies}

\author[0000-0002-5209-5625]{Jean Donet}
\author[0000-0002-0813-5888]{David Wittman}

\affiliation{Department of Physics \& Astronomy, University of California, Davis, CA 
  95616}
\correspondingauthor{David Wittman}
\email{dwittman@physics.ucdavis.edu}

\begin{abstract} 
  Weak gravitational lensing shear could be measured far more precisely if information about unlensed attributes of source galaxies were available.  Disk galaxy velocity fields supply such information, at least in principle, with idealized models predicting orders of magnitude more Fisher information when velocity field observations are used to complement images. To test the level at which realistic features of disk galaxies (warps, bars, spiral arms, and other substructure) inject noise or bias into such shear measurements, we fit an idealized disk model, including shear, to unsheared galaxies in the Illustris TNG100 simulation. The inferred shear thus indicates the extent to which unmodeled galaxy features inject
  noise and bias.  We find that $\gamma_+$, the component of shear parallel to the galaxy's first principal axis, is highly biased and noisy because disks violate the assumption of face-on circularity, displaying a range of intrinsic axis ratios ($0.85\pm0.11$).
  The other shear component, $\gamma_\times$, shows little bias and is well-described by a double Gaussian distribution with central core scatter $\sigma_{\text{core}} \approx$ 0.03, with low-amplitude, broad wings. This is the first measurement of the natural noise floor in the proposed velocity field lensing technique. We conclude that the technique will achieve impressive precision gains for measurements of $\gamma_\times$, but little gain for measurements of $\gamma_+$.
  \end{abstract}
\keywords{\href{	
http://astrothesaurus.org/uat/670}{Gravitational lensing (670)}, \href{	
http://astrothesaurus.org/uat/1797}{Weak gravitational lensing (1797)}, \href{	
http://astrothesaurus.org/uat/671}{Gravitational lensing shear (671)}, \href{	
http://astrothesaurus.org/uat/602}{Galaxy kinematics (602)}, \href{http://astrothesaurus.org/uat/1966}{Magnetohydrodynamical simulations (1966)}}

\section{Introduction} \label{sec:intro}

Weak gravitational lensing is a key tool in modern astrophysics
because it enables mapping of gravitational potentials \citep[see][ for
a recent review]{Bartelmann17}. The physical basis for this mapping is
the gravitational deflection of light from a distant source (assumed
here to be a galaxy) as it passes the potential on its way to the
observer.  The deflection itself is not measurable in the absence of
constraints on the true source position, but variations in deflection
angle across the face of a source galaxy manifest as anisotropic
stretching (shearing) of the galaxy image. Measuring shear thus
requires some knowledge of the unlensed shape of the source.  Rather
than attempt to model unlensed shapes on a source-by-source basis,
weak lensing has proceeded on the assumption that unlensed sources
{\it on average} have no preferred direction. Hence shear measurements
are averaged over many source galaxies in a patch of sky. In this
approach, an individual source carries little information because its
unlensed shape is poorly constrained---a fundamental source of
uncertainty called \textit{shape noise}. This approach is well suited to modern
wide-field imagers that collect data on many sources simultaneously,
but limits the angular resolution of shear measurements.

Information about the unlensed condition of a source galaxy could
enable shear constraints along the specific line of sight to that
galaxy, hence enabling shear maps with better angular resolution.
\citet{Blain2002} suggested a method based on the fact that a circular
orbit inclined to the line of sight produces an observed velocity
pattern that is antisymmetric across its apparent minor axis and
symmetric across its apparent major axis.  This symmetry is disturbed
by lensing, hence any observed asymmetry could be attributed to
lensing.  \citet{Morales2006} extended this idea to disks, deriving a
relation between shear and the angle between the principal axes of the
observed velocity and intensity fields.  \cite{deBurghDay2015}
analyzed velocity maps of nearby (hence unlensed) disk galaxies as
well as simulated velocity maps, and found ${\sim}0.01$ shear precision
per source, compared to ${\sim}0.2$ for traditional weak lensing.

However, these approaches constrain only one of two shear
components. The component of shear directed along the unlensed major
axis, which we define as \gp, preserves the symmetry of the velocity
field but changes the apparent axis ratio.  \citet{Huff2013} proposed
to constrain \gp\ as follows: the Tully-Fisher relation
\citep[TFR;][]{TullyFisher} predicts the (3-D) rotation speed; this
supports an inference of the inclination angle $i$ using the measured
line-of-sight speed; constraining $i$ in turn constrains the unlensed
axis ratio of a circular disk viewed at this inclination.  Hence the
observer can compare lensed and unlensed axis ratios to deduce \gp.
\citet[][, hereafter WS21]{WS21} presented a Fisher Information Matrix
(FIM) analysis of this method using an idealized disk model. They
concluded that the shear precision is a function of inclination angle
$i$, with face-on ($i{=}0$) galaxies offering more precision.  They
found that \gp, as well as the symmetry-violating shear component \gx,
could be inferred to a precision of 0.01 or better (at least at
$i\le 25^\circ$) in this idealized model. However, they warned that
their model accounted only for random measurement errors in the
velocity and intensity fields.  Nature may introduce additional
variations, such as intrinsically noncircular disks upsetting the
axis-ratio calculation, or warped disks that violate the assumed
velocity symmetry.  These variations would constitute the equivalent
of shape noise---a fundamental limit on per-galaxy shear precision due
to variance in unlensed galaxy properties.

The velocity-field method, unlike traditional image-based lensing,
involves different physical arguments for \gp\ and \gx, hence we may
expect different noise levels for each component. This may present
challenges in combining information from multiple sources, but such
challenges are best addressed after completely understanding the
information provided by a single galaxy, which is the focus of this
paper. This paper quantifies the noise floors using simulated
galaxies, which enable high-fidelity mock observations from a variety
of viewing angles.  Although simulated galaxies may omit some details
found in nature, modern simulations incorporate a variety of relevant
physical processes and produce realistic distributions of satellites
and neighbors, bulges, warps, and so on. Hence they offer a useful
starting point free of the complications inherent in using noisy data
obtained from a fixed, unknown viewing angle.  Nevertheless, later in
the paper we will discuss how our results compare to observational
work on closely related issues.


The remainder of this paper is organized as follows:
\S\ref{sec:methods} describes our methods;
\S\ref{sec:results}, the results; and
\S\ref{sec:discussion} provides discussion and interpretation.

\section{Methods} \label{sec:methods}

{\it Overview:} we fit the idealized disk model of
WS21, including the shear parameters \gp\ and \gx, to hundreds of unlensed simulated disk galaxies. The distribution of inferred shear parameters directly indicates the noise and bias due to unmodeled galactic structures, which is likely an irreducible source of error for the velocity field method. In this section we describe the model briefly (\S\ref{subsec-methods-model}), then the galaxy selection (\S\ref{subsec-methods-selection}), Tully-Fisher calibration (\S\ref{subsec-methods-TFRcalib}),
mock data preparation (\S\ref{subsec-methods-mockdata}), and parameter estimation (\S\ref{subsec-methods-opt}).



\subsection{Idealized Model}\label{subsec-methods-model}

We briefly recap the WS21 model and refer readers to that paper for
more details. The model assumes an infinitesimally thin disk with no
bulge, viewed at an angle $i$ between the disk axis and the line of
sight.  The intensity field is described by
$I = I_0 \exp(\frac{-2.99R}{r_{80}})$, where $R$ is a radius in polar
coordinates attached to the disk, $r_{80}$ is the radius that
encircles 80\% of the light, and $I_0$ is the central intensity.  The
rotation curve is described by $v = \vmax \arctan \frac{R}{r_0}$
where $r_0$ is a scale length independent of $r_{80}$, and
$\vmax$ is prescribed by the TFR. The parameter $\ATF$
describes $\vmax$ as a fraction of the TFR prediction, with
a prior allowing for scatter at fixed luminosity. WS21 prescribed 4\% scatter, but we adjusted this for Illustris as described below.  Lensing parameters
in WS21 include \gp, \gx, and the magnification $\mu$; we fix $\mu$ at unity
to focus on the ability of galaxy features to mimic shear. 
The galaxy location on the sky and in
velocity space is described by three nuisance parameters $x_0$, $y_0$,
and $v_0$.  Table~\ref{tab-params} summarizes the parameters.

\begin{table*}
\centering
\caption{Model parameters}
\begin{tabular}{ccccc}
\multicolumn{1}{c}{Symbol} & \multicolumn{1}{c}{Typical value}& \multicolumn{1}{c}{Bounds}&
\multicolumn{1}{c}{Unit}  &\multicolumn{1}{c}{Description} \\ \hline
\ATF & 1 &[0,$\infty$)&-& $v_{\rm max}$ as a fraction of the Tully-Fisher prediction\\
$I_0$ & 90 &[0,$\infty$)&- & intensity S/N at center\\
$i$ & varies &[0,90]& deg & inclination angle \\
$\phi_{\rm sky}$& 90&[0,180)&deg& sky position angle of unlensed major axis\\
$r_{0}$ & 4 &[0,$\frac{L}{2}$)\tablenotemark{$*$}& pixel & rotation curve scale length\\
$r_{80}$ & 12.5&[0,$\frac{L}{2}$) &pixel & radius of 80\% encircled light\\
$x_{0}$ & 0 &(-$\frac{L}{2}$,$\frac{L}{2}$)& pixel & center of galaxy in $x$ coordinate\\
$y_{0}$ & 0 &($-\frac{L}{2}$,$\frac{L}{2}$)& pixel & center of galaxy in $y$ coordinate\\
$v_{0}$ & 0 &-& km/s & galaxy systemic radial velocity\\
$\gamma_+$ & 0 &(-1,1)&- &shear parallel to unlensed major axis \\
$\gamma_\times$ & 0 &(-1,1)&-& shear at 45$^\circ$ to unlensed major axis\\
$\mu$ & 1 & fixed &-& magnification \\
\hline
\end{tabular}
\tablenotetext{*}{$L$ refers to the image side length in pixels (typically $\approx50$ pixels)}
\label{tab-params}
\end{table*}

\subsection{Simulated Galaxies from Illustris TNG100} \label{subsec-methods-selection}

\begin{figure}
    \centering
    \includegraphics[trim=0 0 0 -8mm, scale=0.23]{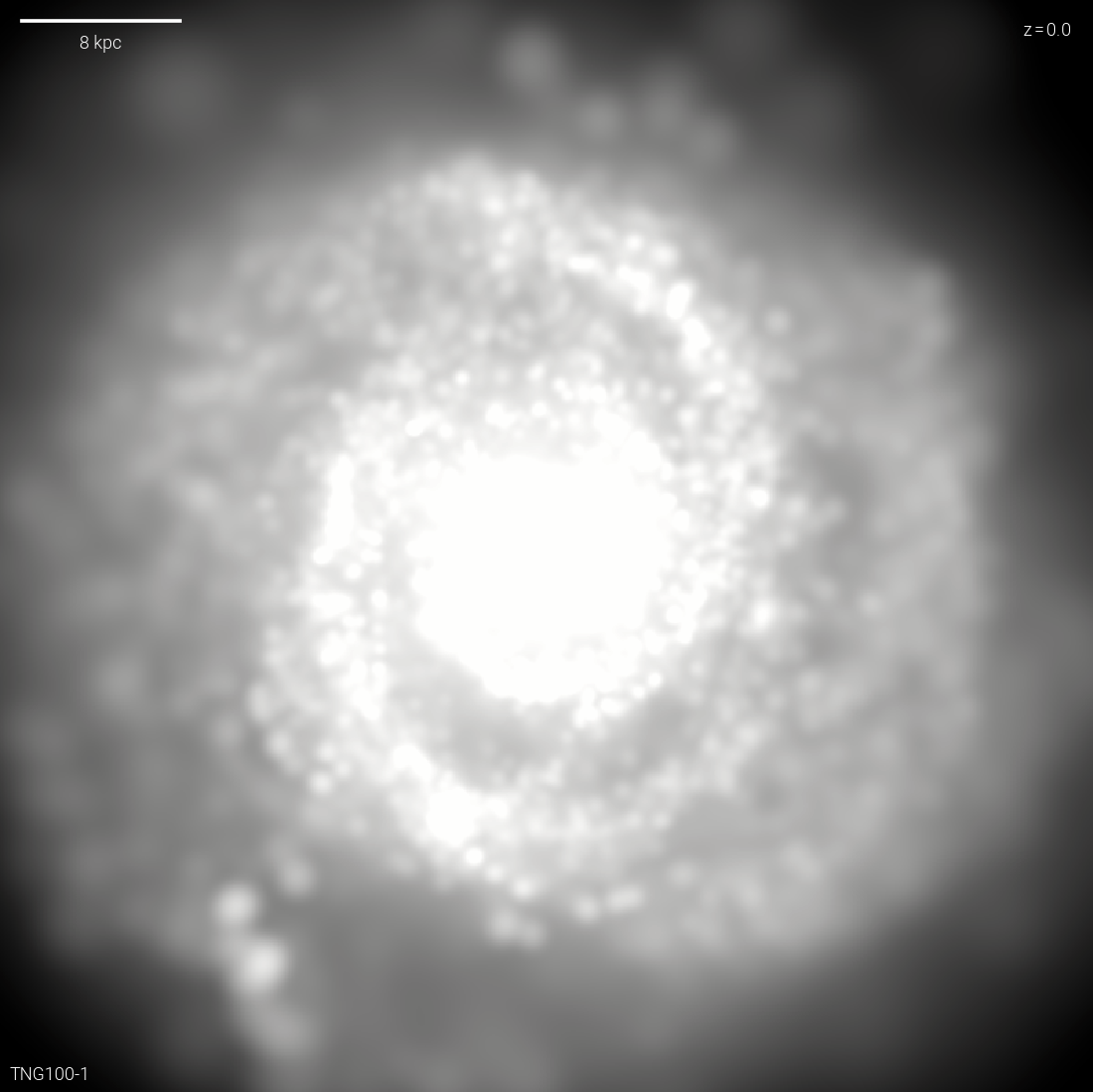}
    \caption{Stellar composite mock-up of subhalo 421835. This
      visualization was generated using Illustris' visualization
      tool. We analyze this subhalo in later figures.}
    \label{fig:galaxyportrait}
\end{figure}

We select galaxies from the $z{=}0$ snapshot of Illustris TNG100-1
\citep{Illustris2019}, a magnetohydrodynamical simulation with a
volume of $110.7^3$ Mpc$^3$ in a Planck2015 cosmology.  The dark
matter and baryon particle masses are $7.5 \times 10^6$ and
$1.4 \times 10^6$ respectively; the softening lengths are 0.74 kpc for
dark matter and a minimum of 0.125 kpc for gas; and the median radius
of star-forming gas cells is 355 pc, with a minimum gas cell size of
14 pc. Illustris TNG includes a wealth of physical effects such as
radiative cooling, stochastic star formation, stellar feedback, and
stellar evolution.  Figure~\ref{fig:galaxyportrait} shows a face-on
view of a disk galaxy in our analysis, demonstrating realistic
features that go far beyond the idealized model of a bulgeless
exponential disk.

To access TNG100 data we use Illustris' web-based Application Programming Interface (API) framework which
allows for a search of all TNG runs and subhalo metadata. This is
done via a specified object id search using the \texttt{requests}
Python module. Our goal is to study massive, rotation-supported galaxies as these are
the galaxies most likely to be targeted in observations if this shear
inference method proves practical.  Furthermore, because Illustris TNG
has many galaxies, we begin with tight selection criteria aimed at
producing a fairly uniform sample of subhalos:

\begin{itemize}
\item the total mass is required to be within 
10\% of 1.4 $\times 10^{12}$ M$_{\bigodot}$
\item $\geq$40\% of the stellar mass is required to be in the disk and
  $\leq$30\% of the stellar mass is allowed to be in the bulge
\end{itemize}
Disk star particles are defined as those with specific angular momenta
nearly as large as the maximum angular momentum of the 100 particles
with the most similar binding energy; in other words with a
circularity parameter \citep{Abadi03} $\epsilon{>}0.7$.  The fraction of
bulge star particles is quantified as twice the fraction with
$\epsilon{<}0$.

These criteria yield 386 subhalos for the fitting described below.

\subsection{TFR Calibration}\label{subsec-methods-TFRcalib}

We examine the TFR for our selected galaxies to verify that the (3-D) rotation speed can be predicted with reasonably small scatter. This is necessary to support inference of $i$ by comparison with the line-of-sight rotation speed; $i$ in turn predicts the unlensed axis ratio of a circular disk, which is compared with the observed axis ratio to infer \gp. 
Our tests do not hinge on reproducing the actual TFR observed in nature. 

\begin{figure}
\centering
    \includegraphics[trim= 0.5cm 0 0 0,scale=0.59]{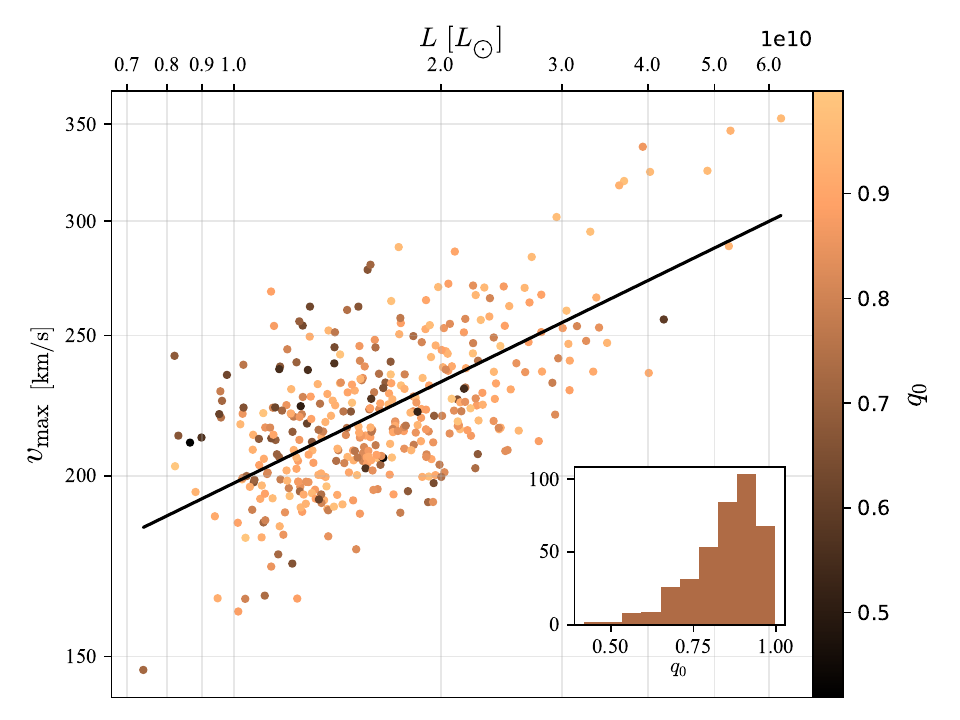}
    \caption{Tully-Fisher relation for the selected halos. The best-fit relation is $L\propto\vmax^{4.3}$ with a scatter of 22 km/s or about 10\%.}\label{fig:TFR}
\end{figure}

We calibrate the TFR on the selected galaxies by using two metadata attributes precomputed by Illustris: the K-band luminosity and the maximum circular speed.  Figure~\ref{fig:TFR} shows these two quantities for each galaxy as well as the best-fit relation $L\propto\vmax^{4.3}$. The residual rms scatter about this relation is 22 km/s or about 10\%.  This is larger than the fiducial model of WS21, but in keeping with the fact that we are using K-band luminosity rather than a baryonic TFR that would require more mock photometry. According to WS21, scatter in the TFR at the 10\% level is not a limiting factor on \gp\ inference, so we proceed with the K-band TFR. With this scatter, the prior on \ATF\ in our fitting procedure becomes $P(\ATF) \propto \exp -\frac{1}{2}(\frac{\ATF-1}{\sigma_{A_{\rm TF}}})^2$ where $\sigma_{A_{\rm TF}}=0.1$.

\subsection{Mock Data Preparation}\label{subsec-methods-mockdata}

For each selected subhalo, we calculate the moment of inertia tensor
of the star particles, and diagonalize the tensor to yield three
principal axes $A\ge B\ge C$ (we use upper-case letters for intrinsic
quantities, reserving lower-case $a$ and $b$ for the apparent semimajor and semiminor axes).  
According to \citet{Franx1992}, intrinsic disk ellipticity may account for a large part of the scatter in the TFR, hence we color-coded the points in Figure~\ref{fig:TFR} by the intrinsic axis ratio $q_0 \equiv \frac{B}{A}$. Although $q_0$ does not seem to correlate with the scatter, the range of $q_0$ highlights the extent to which disks are not circular when viewed face-on; this will become important as follows.
We choose the first principal axis to
coincide with the $x$ axis of a mock detector, and the second
principal axis with the $y$ axis. For views other than face-on, the
second principal axis is foreshortened by a factor of $\cos i$.  This
forces any intrinsic noncircularity to manifest as an inference of
$\gp>0$, which stretches the $x$ axis and shrinks the $y$ axis.

Next we compute the $K$-band intensity field and the line-of-sight
stellar velocity field. We choose a grid with 0.25 kpc/pixel,
extending to $\frac{2}{3}$ of the
\texttt{SubhaloStellarPhotometricsRad} attribute, which is the radius
defined by a surface brightness of 20.7 mag arcsec$^{-2}$ in the $K$
band.  This yields grids with ${\approx}50\times50$ pixels.  The velocity of a pixel is taken to be the mean velocity of the star particles along that line of sight, while the intensity is the sum of the luminosities of those particles. Hence satellites, for example, will cause local departures from a smooth velocity field. We do not model extinction, hence satellites on the far side of the galaxy may be more visible than in real data; dust lanes will not appear but because these are mock K-band observations dust lanes would have been minimal in any case. We assign
uncertainties to each pixel using the WS21 recipe, which is meant to
emulate achievably high signal-to-noise data as follows.  The intensity field
uncertainty is the same in each pixel, and is set to $I_0/90$. The
velocity field uncertainty is set to 10 km/s at the center and scales
inversely with the square root of the local intensity.

\subsection{Least-Squares Optimization}\label{subsec-methods-opt}

We construct a data vector consisting of a
concatenated list of intensity and velocity pixel values, and use an off-the shelf optimizer to find the best-fit model. 
Many of the parameters in our model are bounded by physical or geometric arguments. For example $i$ is defined
on [0,90$^\circ$], $\phi_{\rm sky}$ is defined on [0,180$^\circ$), and the shear components are defined on (-1,1); see Table~\ref{tab-params} for a full list of bounds. To enforce these bounds while efficiently searching for the best-fit model 
we use the Python function
\texttt{scipy.optimize.least\_squares} which implements the ``Trust
Reflective Region'' method \citep{trf99} for dealing with boundaries. 
Within the boundaries, \texttt{scipy.optimize.least\_squares} does not directly support the use of priors. Therefore, we implement the TFR prior indirectly as follows.  \texttt{scipy.optimize.least\_squares} requires a user-defined function that returns a vector of residuals, which it then internally squares and sums to obtain a $\chi^2$ value. Because
$\chi^2=-2\ln\mathcal{L}$ where $\mathcal{L}$ is the likelihood, the \ATF\ prior in \S\ref{subsec-methods-TFRcalib} becomes simply $(\frac{\ATF-1}{\sigma_{A_{\rm TF}}})^2$ in 
$\chi^2$ units. Hence we concatenate one element to the data array with the value $\frac{\ATF-1}{\sigma_{A_{\rm TF}}}$ which is then, internally to \texttt{scipy.optimize.least\_squares}, squared to obtain the $\Delta\chi^2$ corresponding to the prior.

As a test of the code, we fitted mock data generated by the
idealized model. The fitted parameters matched the input parameters to
within one part in $10^6$.  
\vspace{1.5mm}
\section{Results}\label{sec:results}

\begin{figure*}
\centering
    \includegraphics[ width=\textwidth]{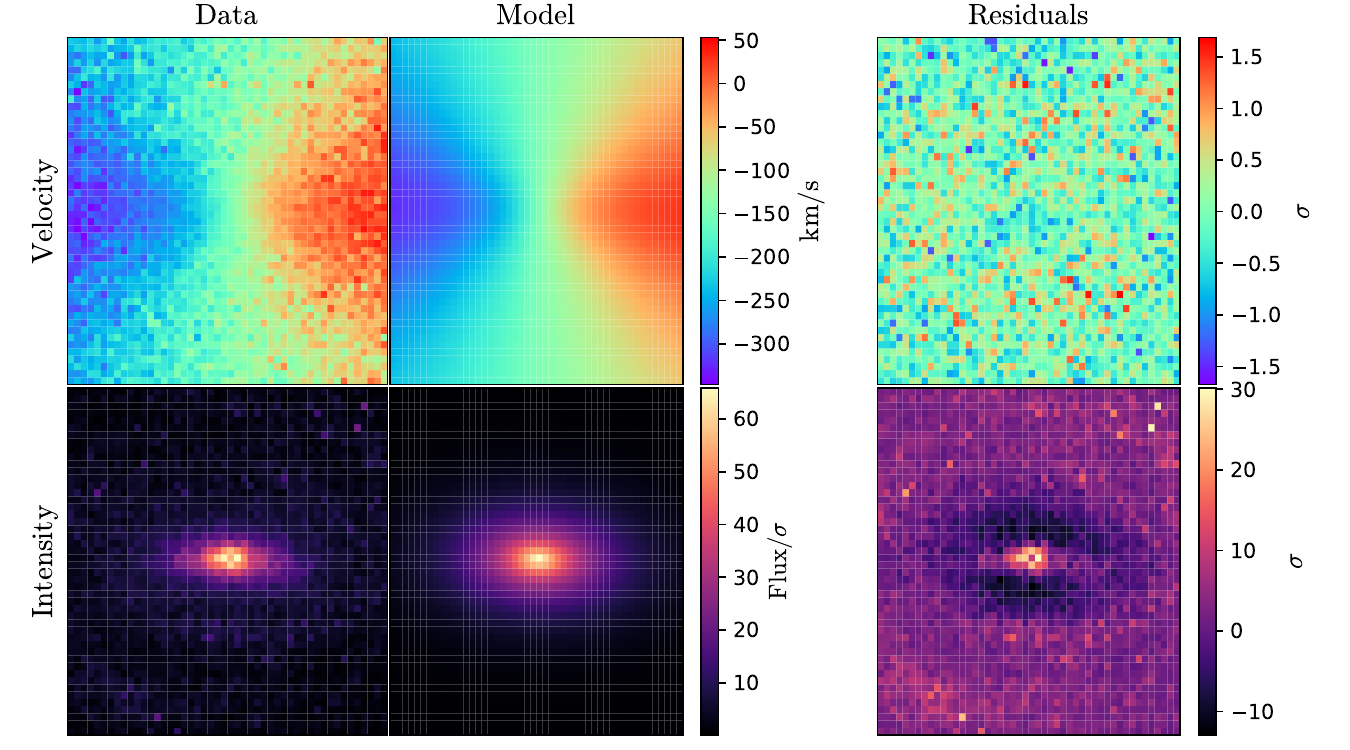}
    \caption{An illustrative TNG100 subhalo (421835, at $z=0$) viewed at $i{=}60^\circ$ from face-on. The top (bottom) row shows the velocity (intensity) field, while the left, middle, and right columns show the mock data, model, and residuals (in units of the uncertainty on each pixel) respectively. The idealized model does not include a bar nor spiral arms, hence these features appear in the residuals.} \label{fig:examplemosaic}
\end{figure*}

We present one example in detail in Figure~\ref{fig:examplemosaic}. The left column contains the data from TNG100, the middle column contains the least-squares optimized model, and the residual of the difference is displayed in the right column. The rows contain the velocity and intensity data and model fields as labelled.  The ``observed'' intensity field has substructure near its center, which violates the idealized model of an axisymmetric disk, but there is little corresponding substructure in the velocity field. The key physical distinction between data and model here is that our model assumes a thin, planar, axisymmetric disk while the mock data from IllustrisTNG assume none of these things. Hence, any shear found by our model fitting may be attributed to galaxy features that violate this model, such as the apparent bar and spiral arms in the lower right panel of Figure~\ref{fig:examplemosaic}. The purpose of this paper is to quantify the extent to which these features would contaminate \gp\ and/or \gx\ measurements using the velocity-field method as implemented by this model.

The extent to which these nonaxisymmetric features result in spuriously inferred shear may vary with viewing angle as well as from galaxy to galaxy. Hence, we 
repeat for each of the 386 galaxies viewed at a series of inclination angles (5, 20, 40, and 60 degrees).  
\vspace{1.5mm}
\subsection{Inferred shear}

\begin{figure}
\centering
    \includegraphics[trim=0.9cm 0 0 0 ,scale=0.46]{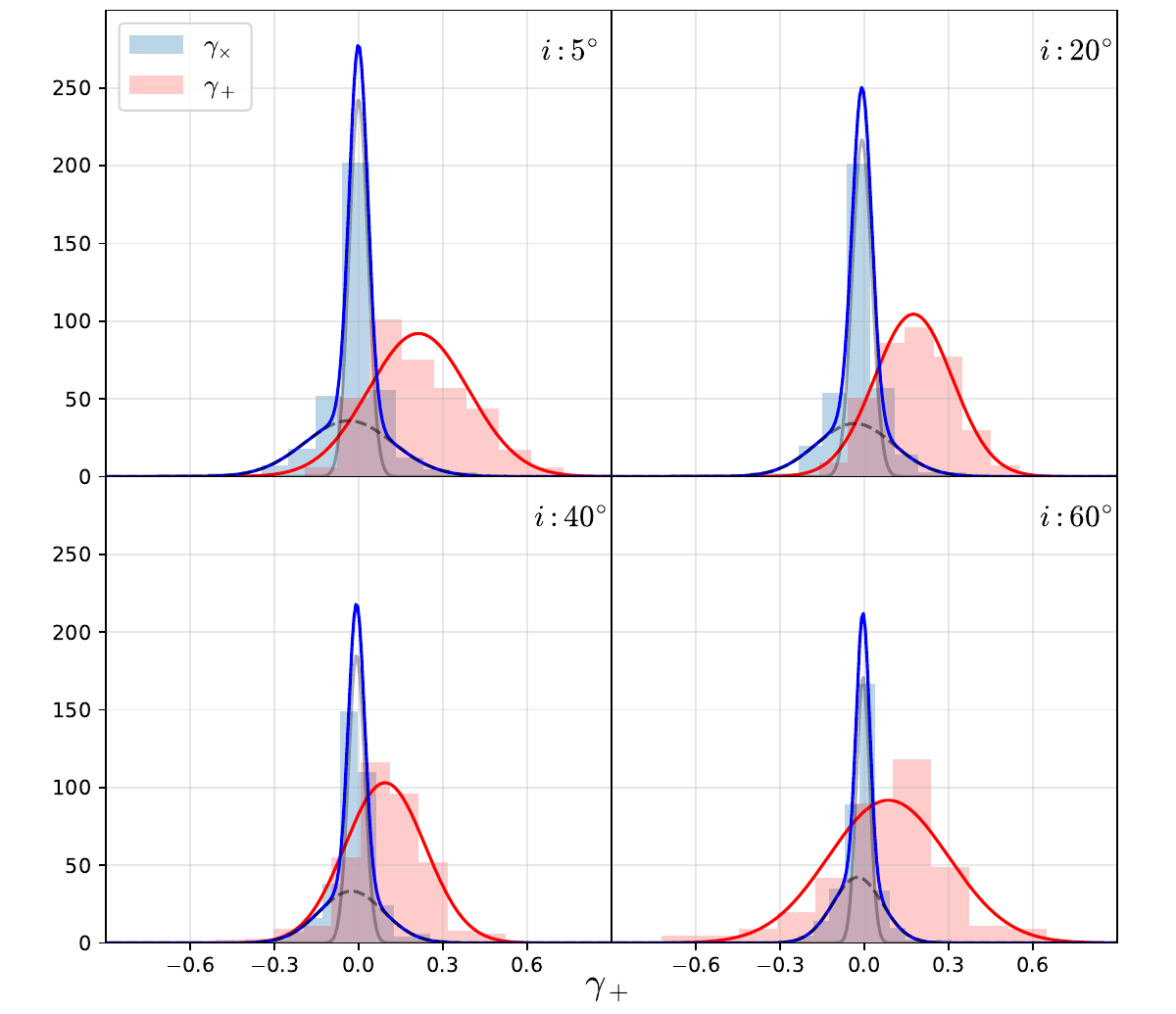}
    \caption{Histograms for inferred $\gamma_+$ and $\gamma_\times$ with overlaid best fit functions. We use a Gaussian mixture model to determine the statistics of the distribution. A Bayesian Information Criterion analysis was conducted to determine the optimal number of model components. $\gamma_+$ is best fit in all cases by a single Gaussian, while $\gamma_\times$ is best fit by a double Gaussian.}
    \label{fig:shearhists}
\end{figure}

Figure~\ref{fig:shearhists} displays histograms of the inferred shear, broken down by component and by true inclination angle. Table~\ref{tab-stats} lists the mean and standard deviation of each of these distributions.  Some of the distributions are non-Gaussian, so for a more complete description we tested Gaussian mixture models (GMM) with a range of Gaussian components and used the Bayesian Information Criterion to determine the preferred number of components in each case. The result is that $\gamma_+$ is best fit at each inclination by a single Gaussian, while $\gamma_\times$ is best fit at each inclination by a double Gaussian.  The resulting descriptive curves are overplotted in Figure~\ref{fig:shearhists}. We discuss each component in more detail below.

\begin{table}
\vspace{2mm}
\caption{Distribution of inferred shear as function of inclination.}
\small
\centering
\centerline{}
\begin{tabular}{@{\extracolsep{4pt}}cccccc}
 & \multicolumn{2}{c}{\gp} & \multicolumn{3}{c}{\gx}\\
 \cline{2-3} \cline{4-6}
$i$ (deg) & mean & $\sigma$ & mean & $\sigma$&$\sigma_{\rm core}$\\ \hline
5 & 0.213&0.179 &-0.016 & 0.103 & 0.033 \\
20& 0.176&0.138 &-0.018 & 0.089 & 0.036\\
40& 0.096&0.143 &-0.015 & 0.076 & 0.029\\
60& 0.084&0.214 &-0.014 & 0.062 & 0.025\\
\end{tabular}
\label{tab-stats}
\end{table}

\vspace{5mm}

\subsection{$\gamma_+$}
\begin{figure*}
\centering
    \includegraphics[width=\textwidth]{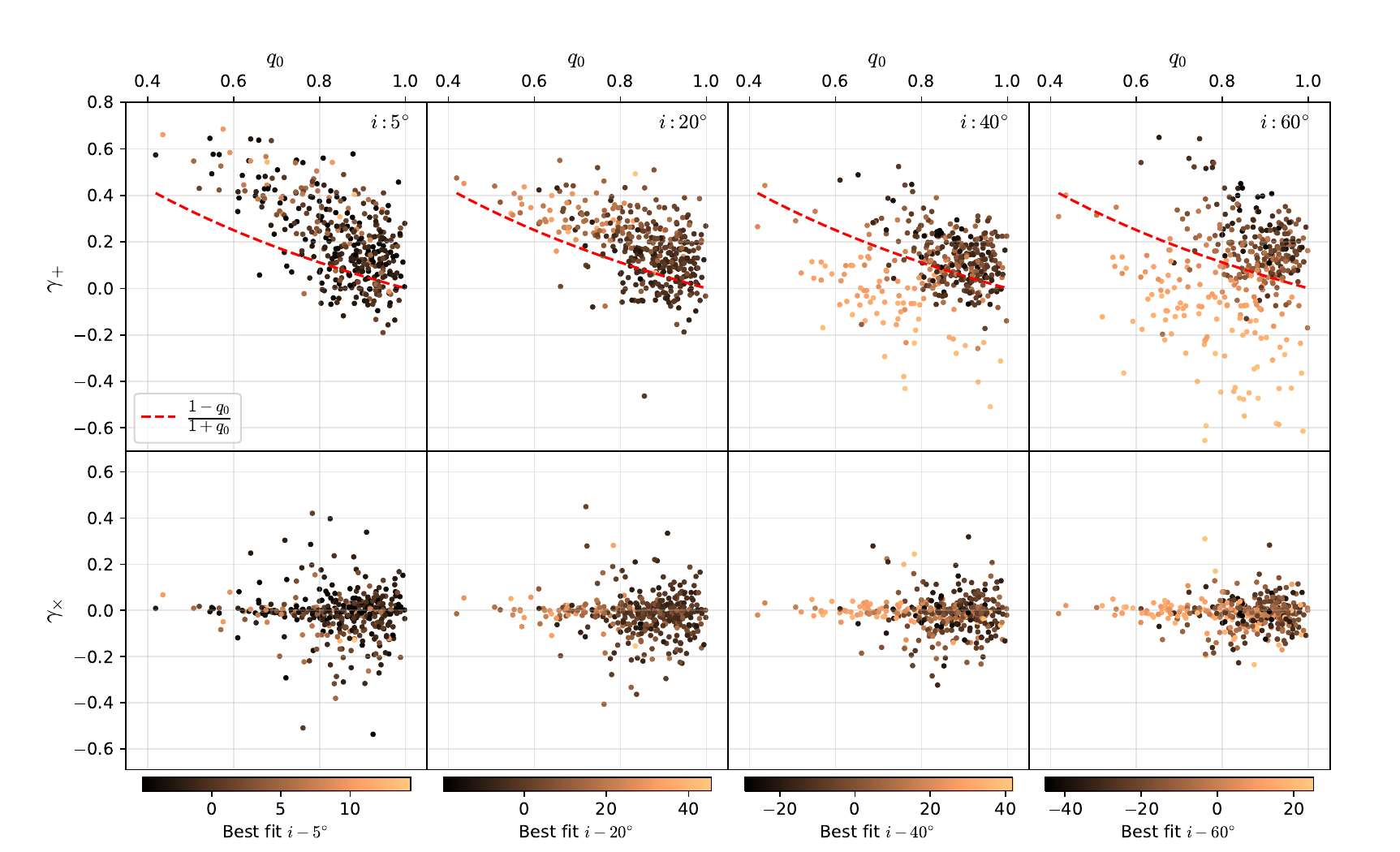}
    \caption{\textit{Top:} scatter plots of inferred $\gamma_+$ vs. $q_0$ for various $i$. We overlay the function $\gamma_+(q_0) = \frac{1-q_0}{1+q_0}$ which describes the expected shear along the major axis arising from the intrinsic axis ratio. We include a colorbar to illustrate the difference between the inferred $i$ and the preset value, indicated at the top right of each panel. \textit{Bottom:} the plots are repeated for \gx. The dashed curve is omitted because $q_0$ is not expected to affect \gx.}
    \label{fig:inferredshear}
\end{figure*}

According to Figure~\ref{fig:shearhists}, \gp\ is biased toward positive values, and shows large scatter (${\approx}0.1-0.2$) at all inclinations.  The sign of the bias suggests that real galaxies depart substantially from the assumption of face-on circularity. We arranged all mock observations such that the true first principal axis $A$ was aligned with the detector's $x$ axis; hence if the second principal axis $B$ is smaller, the galaxy appears as an ellipse with apparent major axis aligned with $x$, even when face-on.  The model has no way to account for this other than introducing $\gp{>}0$, which stretches the apparent $x$ axis of the assumed circular disk. To confirm this hypothesis we plot inferred $\gamma_+$ against the intrinsic axis ratio $\frac{B}{A}\equiv q_0$ in Figure~\ref{fig:inferredshear}. The inverse relation between \gp\ and $q_0$ is clear. We also plot the predicted relationship $\gp = \frac{1-q_0}{1+q_0}$ which follows from inverting the well-known relationship 
$\frac{b}{a} = \frac{1-\gamma}{1+\gamma}$ where $\frac{b}{a}$ is the {\it apparent} axis ratio of a sheared circular disk.

The variation in face-on axis ratio explains much but not all of the spuriously inferred \gp: the inferred \gp\ is generally larger than predicted by the $q_0$ argument, particularly when viewed near face-on. One complicating factor is the anticorrelation with $i$: underestimating $i$ lowers the {\it apparent} axis ratio prediction, which can compensate for overestimating \gp. This trend is revealed by the color gradient in the $i=40^\circ$ and $i=60^\circ$ panels of Figure~\ref{fig:inferredshear}. It is not clear why \gp\ remains overestimated in other panels of this figure where the trend with $i$ is not strong.

Figure~\ref{fig:inferredshear} also demonstrates that constraints on \gp\ degrade as targets become more edge-on. This was predicted by the Fisher forecasts of WS21, and can be understood in terms of a uniform (face-on) velocity field allowing very little wiggle room in $i$. 

\subsection{$\gamma_{\times}$}
As illustrated in Figure~\ref{fig:shearhists}, $\gamma_\times$ performs significantly better than \gp\, with a narrow peak at values close to the expected value of $\gamma_\times {=} 0$. At each inclination the distribution of $\gamma_\times$ is 
best described by a GMM of degree $n {=} 2$, which captures a sharp central peak plus low-amplitude broad wings. To compare with the Fisher forecast of WS21 we examine the width of the central peaks, listed in Table~\ref{tab-params}. The \gx\ precision is several times worse than forecast, and the least favorable viewing angle is near face-on rather than near edge-on as forecast. There is no correlation between \gx\ and other parameters such as $q_0$, $i$, or \gp, so we conclude that the effects discussed in the preceding subsection are not responsible for the greater than expected scatter.

The bias in \gx, although small, is statistically significant (the typical standard error in the mean in each row in Table~\ref{tab-params} is 0.005). This suggests that spiral arms could be responsible for the bias and scatter. Spiral arms will have a greater effect near face-on orientation, and they can cause a bias as follows. We consistently orient the mock data such that the disk is rotating clockwise. Spiral arms that trail \citep[which are the vast majority; ][]{Iye2019} will then trail in an Z-wise rather than S-wise direction on the sky. It is plausible that this could systematically affect the inference of the \gx\ component. Warping of disks could also play a role, but it is more difficult to see why warps would become more important for face-on orientations, or create a systematic effect.

The systematic effect would not be seen in a straightforward average of real observations because any given galaxy is, immutably, seen as S-wise or Z-wise. Z-wise galaxies would have a small bias toward negative \gx\ and S-wise galaxies would have a small bias toward positive \gx.   

\subsection{Additional Tests}

Some readers may find it surprising that galactic disks have such a range of intrinsic axis ratios, hence we spend much of \S\ref{sec:discussion} comparing the IllustrisTNG $q_0$ distribution to other evidence in the literature, with good agreement.  In the remainder of this section, we focus on tests that can be done within IllustrisTNG itself.

First, it is natural to ask whether noncircularity is an artifact of using the star particles as tracers. We repeated the $q_0$ measurements using the mass distributions of the same galaxies. We find that the mass distributions have a more constricted range of $q_0$, but are still markedly noncircular: the mean $q_0$ for mass is 0.92 (with a range of 0.64--1 across 386 galaxies) while the mean for K-band photometry is 0.84 (with a range of 0.42--1).

Second, putting aside the axis ratio question, one may ask whether velocity fields from gas cells perform better. Molecular gas in particular is expected to form a thinner disk, with lower velocity dispersion, than stars. We performed fits using molecular-gas velocity fields, and found that performance was worse. This can be attributed to the patchy nature of molecular gas, which forms clumps around the spiral arms. The low velocity dispersion of molecular gas does not provide a substantial advantage in our mock data because the uncertainty on the mean velocity of each pixel is assumed to be dominated by measurement uncertainty rather than intrinsic dispersion. It remains possible that an alternative modeling procedure could do better with molecular gas velocity fields, but that is beyond the scope of this paper.  We also found that velocity fields from atomic gas did not perform well.

Third, one may ask whether outliers can be identified and rejected based on observable criteria. However, we found little correlation between a poor fit (as determined by $\chi^2$) and spuriously inferred shear. We also visually inspected the 20 galaxies with the largest spurious shear to look for commonalities such as the presence of bars or satellites. We found no unambiguous trends. Bars, for example, may be overrepresented in the outliers, but a substantial fraction of outlying galaxies do not have bars. We also looked for a correlation between spurious shear and the presence of nearby galaxies which may induce tidal distortions in the target galaxy. For each IllustrisTNG galaxy of mass $M$ and distance $r$ from the target galaxy, we tabulate the tidal acceleration at the location of the target galaxy, $\frac{M}{r^3}$; we then keep the largest tidal acceleration experience by each target galaxy. We find no correlation between this tidal acceleration and spuriously inferred shear.

\section{Discussion \& Summary}\label{sec:discussion}

The scatter and bias in \gp\ will be serious obstacles in attempting to make the velocity-field method competitive with traditional image-based weak lensing, at least for this component of shear. Image-based weak lensing yields a slightly larger shear uncertainty per galaxy (${\sim}0.2$ vs. 0.1--0.2), but it involves far less expensive observations that collect many more source galaxies, and it makes no assumptions about the intrinsic shapes of galaxies. The mediocre \gp\ performance of the velocity-field method is due to scatter in the face-on axis ratio $q_0$, which further causes a bias because galaxies can only scatter in one direction from the assumed value of $q_0=1$. Hence, our conclusion hinges on IllustrisTNG reproducing the $q_0$ distribution in the observed universe, at least qualitatively. The bulk of this section examines the literature on that issue and confirms the conclusion. 

\begin{figure*}
    \centerline{\includegraphics[ width=8.3cm,trim=0 0 0 0]{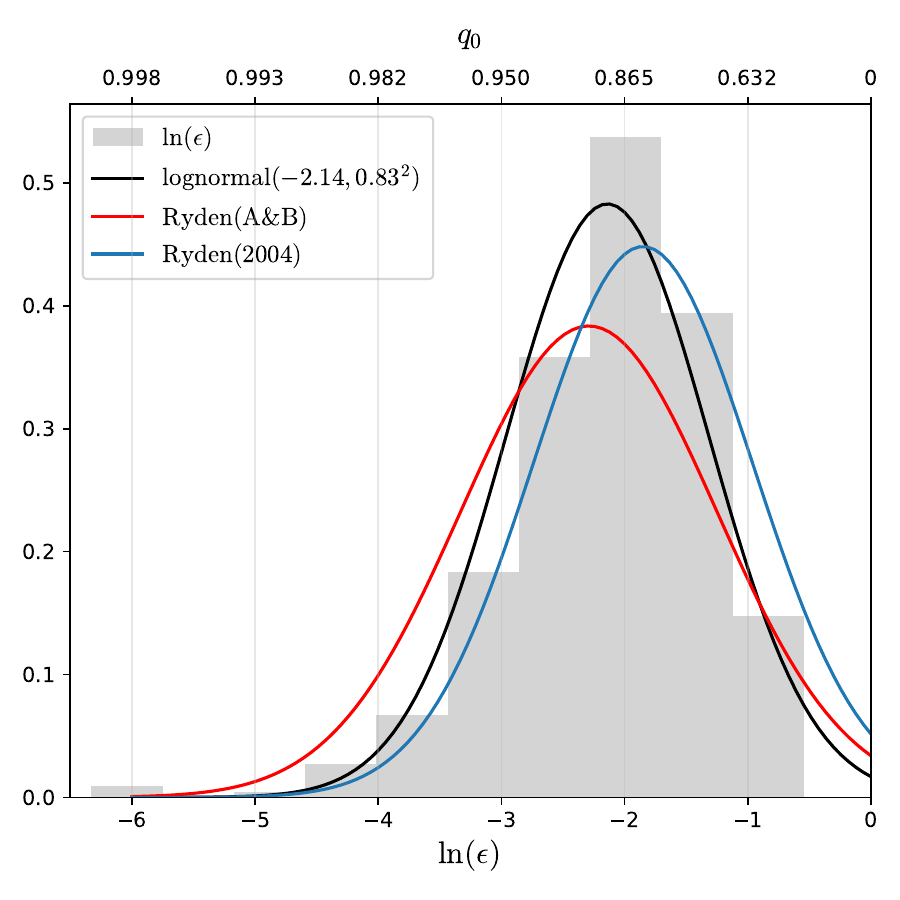}
        \includegraphics[width=8.5cm]{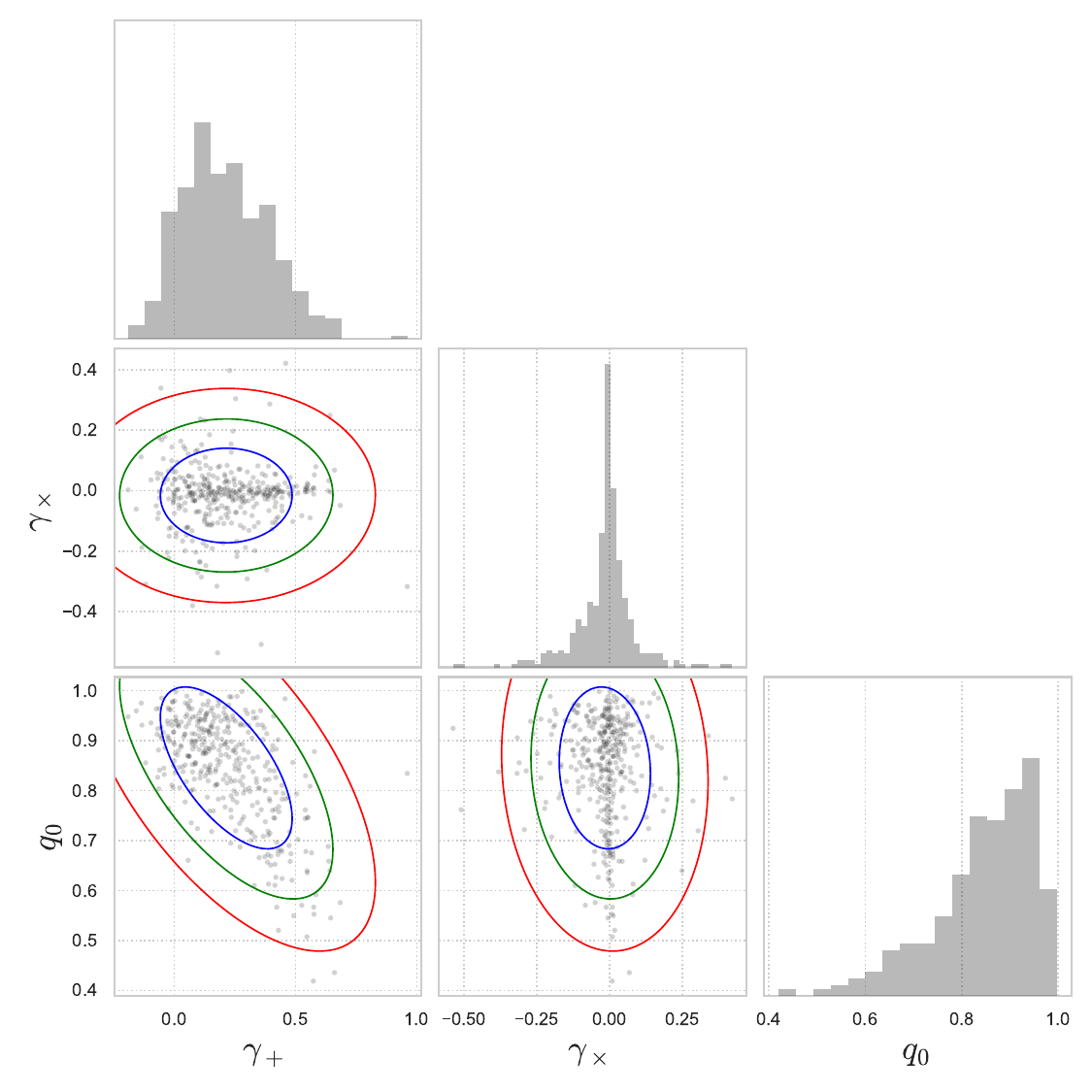}
    }
    \caption{Understanding the scatter in $\gamma_+$: \textit{Left:} Distribution comparison for our sample of 386 subhalos in the $K$ band, Ryden's correction of \citet{AndersenBershady02} in the $I$ band, and \citet{Ryden04} in the $i$ band. Our sample is best described by a lognormal distribution peaking at $\ln(\epsilon){=}-2.14 \pm 0.83$, we include the equivalent axis ratio values at the top for clarity.  \textit{Right:} Corner plot showing that $q_0$ is anticorrelated with $\gamma_+$ but has no correlation with  $\gamma_\times$. Contours show 68.27\%, 95.45\%, and 99.73\% confidence intervals.}
    \label{fig:ryden}
\end{figure*}

We briefly recap two methods of inferring $q_0$ from observations. The
first \citep{Ryden04,Ryden06} finds the distribution of $q_0$ that, when
viewed from random angles, best matches the observed distribution of
apparent axis ratios.  In 12,764 SDSS galaxies \citet{Ryden04} found a
median axis ratio of 0.85 and a scatter ${\approx}0.1$ in $i$ band. Figure~\ref{fig:ryden} shows that her best-fit distribution (blue curve) is quite similar to the distribution we extracted from Illustris (black curve). \citet{Ryden06}
extended this type of analysis to include $K$-band data from 2MASS, as
well as a breakdown into early and late spirals. She
found a strong trend in median $q_0$ with Hubble type: from ${\approx}0.98$ for late spirals to ${\approx}0.7$ for early spirals. This is again comparable to the range in Illustris.  If $q_0$ is indeed a strong function of Hubble type, a possible workaround for velocity-field shear inference is to adopt a prior on 
$q_0$ for each observed galaxy based on its Hubble type.  However, there remains substantial scatter in $q_0$ at fixed Hubble type so this prior would remove only some of the scatter in inferred \gp.

The second method \citep{AndersenBershady01,AndersenBershady02} is
capable of inferring $q_0$ on a per-galaxy basis.
\citet{AndersenBershady01} used intensity and velocity field data to
compare the sky position angle (PA) of the photometric and kinematic
major axes.  Such a misalignment can be caused by lensing, but not in
the nearby sample studied by \citet{AndersenBershady01}. They
attributed any misalignment to the disk having circular orbits but an
intrinsically photometric ellipticity $\epsilon\equiv1-q_0$. They examined seven galaxies and found $\epsilon$ ranging from 0.02--0.20.
\citet{AndersenBershady02} examined 28 galaxies and found $\epsilon=0.06^{+0.064}_{-0.031}$;
\citet{Ryden04} corrected this for selection effects and derived a distribution shown as the red curve in Figure~\ref{fig:ryden}. Our Illustris distribution differs from the two Ryden results by less than those two results differ from each other. We conclude that Illustris is accurately reproducing the intrinsic noncircularity of disk galaxies.

We note that velocity fields are typically measured for interstellar gas---e.g. 
\citet{AndersenBershady01} and \citet{AndersenBershady02} used the H$\alpha$ emission line---while intensity fields are typically measured for stars, e.g.
these references used broadband $R$ and $I$ images. Any misalignment between gas and stellar velocity fields would contribute to the ellipticity inferred by these authors, but would not manifest in our results because we used the stellar velocity field, nor in Ryden's results which are based entirely on photometry. The consistency of these three results suggests that gas-star misalignments are a subdominant effect. 

Although the details may vary with galaxy selection and wavelength,
this body of work consistently implies that disks are not
intrinsically circular, which leads to a bias of $\approx0.1$ on the
\gp\ inference. Even if the mean effect of intrinsic ellipticity could
be removed from the shear inference---an open question, considering
the unknown PA of the intrinsic ellipticity---there remains a scatter
of the same order, possibly reducible only through careful selection
of very late types if the \citet{Ryden06} results are correct.  This
casts serious doubt on the viability of the kinematic method for
inferring \gp.

This issue does not affect $\gamma_\times$, where we see promising precision gains compared to traditional weak lensing. The \gx\ scatter has a tight central core with $\sigma_{\text{core}}{\approx}0.03$, dramatically 
smaller than the ${\sim}0.2$ from image-based weak lensing. 
The scatter is larger than predicted by the idealized model of WS21, as expected from galaxies with realistic features such as spiral arms and warps; this work documents for the first time how much scatter is contributed by such features. 

It is instructive to compare our \gx\ results with those of \citet{deBurghDay2015}, whose inference method involves a \textit{nonparametric} disk model created by reflecting the data itself. In their method, the inferred value of \gx\ is that which minimizes the asymmetry when unlensed; \gp\ is not constrained. They tested this method on idealized disks similar to those created by our parametric model, and found uncertainties ${\approx}0.005$ with mock data similar in quality to ours. This is in line with the WS21 Fisher prediction, and suggests that the ${\gtrsim}0.03$ scatter we see is largely due to galaxy features not captured by our parametric model. Surprisingly, when \citet{deBurghDay2015} applied this method to two actual (nearby, hence unlensed) galaxies, the \gx\ precision was much tighter, ${\approx}0.001$. It is unclear how higher precision was obtained from less idealized data.

It is possible that, for \gx\ inference, nonparametric galaxy models are more robust against features such as bars, warps, and spiral arms. This suggestion is tentatively supported by the small \gx\ bias in our mock data, which were prepared with all galaxies oriented Z-wise. Future work on \gx\ could include further testing of nonparametric models and perhaps comparison with model extensions for bars, warps, and arms. Nevertheless, while the inference of \gx\ from the velocity field could benefit from further modeling work, it clearly enables substantial precision gains over traditional image-based weak lensing. We also note that additional model parameters are unlikely to ameliorate the \gp\ issue, which is caused by the degeneracy of the two existing parameters \gp\ and $q_0$; the latter parameter was fixed in our tests but the degeneracy is clear. Hence, practical applications of velocity-field lensing will require strategies that take advantage of precise measurements of \textit{one} component of shear defined relative to each source galaxy's sky orientation. For mass mapping this may require using pairs of differently oriented galaxies \citep{Blain2002}; however mass \textit{models} can naturally be fit to the data regardless of widely varying precision on the two shear components along a given line of sight. For cosmic shear, cosmological models can predict the correlations of randomly selected shear components just as they can predict the auto- and cross-correlations of both components, although the reduction in model discrimination power may be substantial.

Finally, we note that this method and traditional weak lensing will have different sources of \textit{systematic} error, which could enable cross-checks between the methods. Initial applications of the velocity-field method will presumably focus on a small number of well-resolved sources for which PSF modeling will be a subdominant source of error. Traditional weak lensing, in contrast, uses large numbers of poorly resolved galaxies for which PSF modeling is the dominant systematic error. Measuring even one component of shear more precisely along selected lines of sight could potentially test the calibration of traditional weak lensing surveys.

\begin{acknowledgments}
We thank Dylan Nelson for patiently answering many questions about
Illustris-TNG, and we also thank the anonymous referee. This work was supported by NSF grant AST-1911138.
\end{acknowledgments}


\bibliography{main}{}
\bibliographystyle{aasjournal}



\end{document}